\documentclass[aps,pre,onecolumn]{revtex4}%
\usepackage[english]{babel}
\usepackage{amsmath}
\usepackage{color}
\usepackage{epsfig}
\usepackage{graphicx}
\usepackage{bm}
\usepackage{times,amsmath,amssymb}
\usepackage{subfigure}
\usepackage{amsfonts}
\usepackage{amssymb}%
\begin{document}
\title{Alternation of sign of magnetization current in driven $XXZ$ chains with
twisted $XY$ boundary gradients}
\author{V. Popkov$^{1,2}$}
\affiliation{$^{1}$ Dipartimento di Fisica e Astronomia, Universit\`a di Firenze, via G.
Sansone 1, 50019 Sesto Fiorentino, Italy}
\affiliation{$^{2}$ Max Planck Institute for Complex Systems, N\"othnitzer Stra{\ss }e 38,
01187 Dresden, Germany}

\begin{abstract}
We investigate an open $XXZ$ spin $1/2$ chain driven out of equilibrium by
coupling with boundary reservoirs targeting different spin orientations in
$XY$ plane. Symmetries of the model are revealed which appear to be different
for spin chains of odd and even sizes. As a result, spin current is found to
alternate with chain length, ruling out the possibility of ballistic
transport. Heat transport is switched off completely by virtue of another
global symmetry. Further, we investigate the model numerically and
analytically. At strong coupling, we find exact nonequilibrium steady state
using a perturbation theory. The state is determined by solving secular
conditions which guarantee self-consistency of the perturbative expansion. We
find nontrivial dependence of the magnetization current on the spin chain
anisotropy $\Delta$ in the critical region $\left\vert \Delta\right\vert <1$,
and a phenomenon of tripling of the twisting angle along the chain for narrow
lacunes of $\Delta$.

\end{abstract}
\date{\today }
\maketitle

\section{Introduction}

Manipulating of quantum systems which consist of a few quantum
dots or quantum bits forms a basis of a functioning of any quantum
computing device. Recent experimental advances allow to assemble
and manipulate nanomagnets consisting of just a few  atoms and
perform measurements on nanostructures with an atomic resolution.
However, a theoretical understanding of microscopic quantum
systems out of equilibrium (e.g. under constant pumping or
continuous measurement by a quantum probe) is far from being
complete, apart from simplest cases like a single two-level system
or a quantum harmonic oscillator under external pumping or in
contact with a reservoir.

Coupling of a quantum system with an environment (or with a measuring
apparatus) is described under standard assumptions \cite{Petruccione},
\cite{PlenioJumps} in the framework of a Lindblad Master equation (LME) for a
reduced density matrix, where a unitary evolving part is complemented with a
Lindblad dissipative action. Under the LME dynamics, a system with gradients
evolves towards a nonequilibrium steady state (NESS) with currents.

The Heisenberg model of interacting spins is an oldest many-body quantum model
\ of a strongly correlated systems, important both from a theoretical and an
experimental point of view. Energy and magnetization transport in the $XXZ$
Heisenberg spin chain is extensively studied by different methods, see
\cite{review1, zotos} for reviews, including Bethe Ansatz \cite{shastry,
zotos99, KlumperLectNotes2004}, bosonization \cite{Bosonisation},
\cite{Bosonisation1}, Lagrange multipliers \cite{Antal97-98}, exact
diagonalization \cite{zotos96, andrei98}, Lanczos method \cite{lanczos},
quantum Monte Carlo \cite{alvarez}, DMRG \cite{SchollwoeckRev05,Gobe05}.
Investigation of nonequilibrium spin chains from the point of view of
Markovian Master equations contributes to the "equilibrium" studies. In quasi
one-dimensional spin chain materials like SrCuO$_{2}$ many transport
characteristics are measurable experimentally \cite{HessBallistic2010}.

Investigation of NESS in driven spin chains is hindered by enormous technical
difficulties due to an exponentially growing Hilbert space, although
time-dependent DMRG method allows to reach somewhat larger sizes
\cite{ZnidaricJStat2010_2siteLindblad}. In many studies the NESS\ are produced
by coupling a spin chain at the edges to boundary baths of polarization
aligning boundary spins along the anisotropy axis $Z$, so that the first and
the last spin tend to be antiparallel \cite{remark_antiparallel}. For this
particular choice of boundary baths, a substantial progress in understanding
the properties of a driven $XXZ$ model have been recently achieved
\cite{Pros08,ZunkovichJStat2010ExactXY,BenentiPRB2009,ProsenZnidaric,ProsenExact2011,
JesenkoZnidaric2011,znidaricprl2011,ZnidaricJStat2010,ZnidaricJPhysA,ClarkPriorMPA2010,ProsenNY2010}%
, including an exact solution of the problem in case of strong boundary
driving \cite{ProsenExact2011}, see also \cite{JesenkoZnidaric2011} and
references therein. Note that within this boundary setup, one has the
transport of a $z$-component of magnetization $j^{z}$, caused by boundary
gradients along the $z$-axis, thus probing diagonal elements of a diffusion
matrix and a conductivity tensor.

In present paper we carry out a systematic study of nonequilibrium steady
states in driven $XXZ$ spin chain resulting from boundary gradients in the
$XY$ direction, transverse to the anisotropy axis. In the context of spin and
energy transport and for appropriate limits, this would correspond to probing
the nondiagonal elements of the diffusion and conductivity tensors. To this
end, we bring an $XXZ$ spin chain in contact with reservoirs aligning the
boundary spins along the $X$-axis at one end and along the $Y$-axis in the
other end, thus trying to impose a twisting angle in $XY$-plane of $\pi/2$
between the first and the last spin. Within the LME formalism, twisting angles
$\alpha=0$ or $\alpha=\pi$ result in a complete suppression of the
magnetization current, see a remark after Eq.(\ref{CurrentNonBallistic}),
making a setup with $\alpha=\pi/2$ the simplest nontrivial one. We derive
general symmetries of the model which indicate that magnetization current
should alternate its sign with system size, for any value of boundary
coupling. On this basis we rule out the possibility of a ballistic spin
current, which is typically observed in integrable $XXZ$ model under other
kinds of perturbations \cite{ZotosBallistic97},\cite{ProsenMasur2011}. We
argue, further, that the phenomenon of the spin current alternation is rather
general and in particularly is not related to the integrability.

To confirm further our findings we investigate the nonequilibrium steady state
of the $XXZ$ model numerically and analytically, for varying amplitudes of
boundary couplings and boundary gradients. We find complete agreement with our
predictions which in particular are confirmed analytically, in the regime of
strong coupling. In this regime, full analytic solution of the Lindblad
equation for large Lindblad amplitudes demonstrates a very nonmonotonic
dependence of the magnetization current on the system size and anisotropy.

The plan of the paper is the following. In sec.\ref{sec::The model} we
introduce the model. Symmetries of the Lindblad master equation are obtained
in
sec.\ref{sec::Symmetries of the Lindblad equation and restrictions on energy and magnetization currents}%
. Using the symmetries general NESS properties for arbitrary system size and
coupling are revealed, including an admissibility of spin and energy currents.
Parity-depending LME symmetries lead to a phenomenon of magnetization current
sign alternation with the system size. In sec.
\ref{sec::Strongly driven XXZ chain : numerical and analytical study} our
predictions are tested further by calculating the NESS\ analytically and
numerically for small system sizes. The cases of $XX$ model and of $XXZ$ model
are treated separately in sec.\ref{subsec::Free fermion point. Arbitrary N}
and sec.\ref{subsec::Arbitrary Delta. Finite size results}. Limits of weak
coupling and of weak driving force are briefly discussed at the end of
sec.\ref{subsec::Generic Gamma and arbitrary driving}. Appendix contain
necessary technical details. In conclusion we summarize our findings.

\section{The model}

\label{sec::The model}

\begin{figure}[ptb]
\begin{center}
\centerline{\scalebox{0.6}{\includegraphics{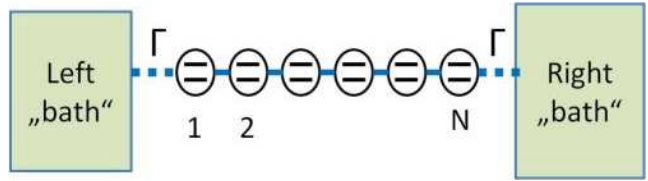}}}
\end{center}

\caption{ Schematic layout of the problem. The chain of two-level systems
coupled at the boundaries to the reservoirs. Dissipation introduced by the
reservoirs is described by a quantum Master equation (\ref{LME}) }%
\label{Fig_LindbladReservoirs}%
\end{figure}

We study an open chain of $N$ quantum spins in contact with boundary
reservoirs. The time evolution of the reduced density matrix $\rho$ is
described by a quantum Master equation in the Lindblad form \cite{Petruccione}%
,\cite{PlenioJumps}, \cite{ClarkPriorMPA2010} (here and below we set $\hbar=1$)%

\begin{equation}
\frac{\partial\rho}{\partial t}=-i\left[  H,\rho\right]  +\Gamma(%
\mathcal{L}%
_{L}[\rho]+%
\mathcal{L}%
_{R}[\rho]), \label{LME}%
\end{equation}
where $H$ is the Hamiltonian of an extended quantum system, and $%
\mathcal{L}%
_{L}[\rho]$ and $%
\mathcal{L}%
_{R}[\rho]$ are Lindblad dissipators acting on spatial edges of the system.
This setting, shown schematically in Fig.\ref{Fig_LindbladReservoirs}, is a
common starting point in studies of nonequilibrium transport, a particular
choice of $%
\mathcal{L}%
_{L}$ and $%
\mathcal{L}%
_{R}$ depending on the application. Often, a choice of the Lindblad action is
operational, i.e. is determined by a requirement to favour a relaxation of a
system or its part towards a target state with given properties, e.g. to a
state with given polarization or temperature
\cite{ZnidaricJStat2010_2siteLindblad}. In this way, one describes an
effective coupling of the system, or a part of it, with a respective
bath-environment. Note that within the quantum protocol of repeated
interactions, \cite{ClarkPriorMPA2010}, \cite{RepeatedInteractionScheme} the
equation (\ref{LME}) appears to describe an exact time evolution provided that
the coupling to reservoir is rescaled appropriately with the time interval
between consecutive interactions of the system with the reservoirs.

We specify the Hamiltonian to describe an open $XXZ$ spin chain with an
anisotropy $\Delta$
\begin{equation}
H=\sum_{k=1}^{N-1}\left(  \sigma_{k}^{x}\sigma_{k+1}^{x}+\sigma_{k}^{y}%
\sigma_{k+1}^{y}+\Delta\sigma_{k}^{z}\sigma_{k+1}^{z}\right)  ,
\label{Hamiltonian}%
\end{equation}
while $%
\mathcal{L}%
_{L}$ and $%
\mathcal{L}%
_{R}$ are Lindblad dissipators favouring a relaxation of boundary spins $k=1$
and $k=N$ towards states described by one-site density matrices $\rho_{L}$ and
$\rho_{R}$, i.e. $%
\mathcal{L}%
_{L}[\rho_{L}]=0$ and $%
\mathcal{L}%
_{R}[\rho_{R}]=0$. General form of the matrices $\rho_{L}$ and $\rho_{R}$ is
$\rho_{L,R}=\frac{I}{2}+\alpha_{L,R}\sigma^{x}+\beta_{L,R}\sigma^{y}%
+\gamma_{L,R}\sigma^{z}$, where $\sigma^{a}$ are Pauli matrices, and
$\alpha,\beta,\gamma$ are real constants satisfying $Tr\rho_{L,R}^{2}\leq1$ or
$\alpha_{L,R}^{2}+\beta_{L,R}^{2}+\gamma_{L,R}^{2}\leq1$. For a general choice
of the Lindblad dissipator $%
\mathcal{L}%
_{L,R}[.]=\sum_{i,j=1}^{3}a_{ij}^{L,R}\left(  \sigma^{i}.\text{ }\sigma
^{j}-\frac{1}{2}\{\text{ }.\text{ },\sigma^{j}\sigma^{i}\}\right)  $, each of
the $2\times2$ matrices $\rho_{L}$ and $\rho_{R}$, if not constant, belongs to
at most one-dimensional manifold \cite{Gorini76}.

We choose boundary reservoirs which tend to align the spin at the left edge
along the $x$ direction and the spin at the right edge along the $y$
direction. Consequently, the reservoirs try to establish a twisting angle of
$\pi/2$ between the first and the last spin in the chain. Such a setting is
achieved by taking the Lindblad action in the form%

\begin{align}%
\mathcal{L}%
_{L}[\rho]  &  =-\frac{1}{2}\sum_{m=1}^{2}\left\{  \rho,W_{m}^{\dag}%
W_{m}\right\}  +\sum_{m=1}^{2}W_{m}\rho W_{m}^{\dag},\label{LindbladAction}\\%
\mathcal{L}%
_{R}[\rho]  &  =-\frac{1}{2}\sum_{m=1}^{2}\left\{  \rho,V_{m}^{\dag}%
V_{m}\right\}  +\sum_{m=1}^{2}V_{m}\rho V_{m}^{\dag}.\nonumber
\end{align}
Here $W_{m}$ and $V_{m}$ are polarization targeting Lindblad operators, which
act on the first and on the last spin respectively,
\begin{align}
W_{1}  &  =\sqrt{\frac{1-\kappa}{2}}(\sigma_{1}^{z}+i\sigma_{1}^{x}%
),\nonumber\\
W_{2}  &  =\sqrt{\frac{1+\kappa}{2}}(\sigma_{1}^{z}-i\sigma_{1}^{x}%
),\nonumber\\
V_{1}  &  =\sqrt{\frac{1+\kappa^{\prime}}{2}}(\sigma_{N}^{y}+i\sigma_{N}%
^{z}),\label{LindbladOperators}\\
V_{2}  &  =\sqrt{\frac{1-\kappa^{\prime}}{2}}(\sigma_{N}^{y}-i\sigma_{N}%
^{z}).\nonumber
\end{align}
In absence of the unitary term in (\ref{LME}) the boundary spins relax with a
characteristic time $\Gamma^{-1}$ \cite{BoundaryRelaxationTimes} to specific
states described via \textit{one-site} density matrices $\rho_{L}$ and
$\rho_{R}$, satisfying $%
\mathcal{L}%
_{L}[\rho_{L}]=0$ and $%
\mathcal{L}%
_{R}[\rho_{R}]=0$, where
\begin{align}
\rho_{L}  &  =\frac{1}{2}\left(  I-\kappa\sigma_{1}^{y}\right) \label{RoL}\\
\rho_{R}  &  =\frac{1}{2}\left(  I+\kappa^{\prime}\sigma_{N}^{x}\right)  .
\label{RoR}%
\end{align}
From the definition of a one-site density matrix $\rho_{\text{one-site}}%
=\frac{1}{2}\left(  I+%
{\textstyle\sum}
\langle\sigma^{\alpha}\rangle\sigma^{\alpha}\right)  $, we see that the
Lindblad superoperators $%
\mathcal{L}%
_{L}$ and $%
\mathcal{L}%
_{R}$ indeed try to impose a twisting angle of $\pi/2$ in $XY$ plane between
the first and the last spin \cite{TwistingRemark}. The twisting gradient
drives the system in a steady-state with currents.\ In the following we
restrict to a symmetric choice%
\begin{equation}
0\leq\kappa=\kappa^{\prime}\leq1. \label{SymmetryU-Uprime}%
\end{equation}
The parameter $\kappa$ determines the amplitude of the gradient between the
left and right boundary, and therefore plays a fundamentally different role
from the coupling strength $\Gamma$. The limits $\kappa=1$ and $\kappa\ll1$
will be referred to as the strong driving and weak driving case respectively.
The two limits describe very different physical situations as exemplified in
other NESS\ studies.

The reason for choosing the Lindblad dissipators which impose a twisting
gradient in $XY$-plane perpendicular to the Heisenberg chain anisotropy is
two-fold. Firstly, our choice is "orthogonal" and therefore complementary to
usually considered boundary setup \cite{Pros08}-\cite{JesenkoZnidaric2011}%
,\cite{znidaricprl2011}-\cite{ZnidaricJPhysA}, where alignment of spins at the
boundary is parallel to the anisotropy axis $Z$. Our system is not mappable to
a quadratic fermionic system, and is therefore not amenable to a analysis via
a canonical quantization in the Fock space of operators \cite{Pros08}%
,\cite{ZunkovichJStat2010ExactXY}, which can be used for usual boundary setup.
Secondly, one realizes that imposing both $X$- and $Y$- boundary gradients is
necessary to create a nonvanishing stationary $Z$-magnetization current. A
boundary setup with parallel or antiparallel boundary spins, aligned in the
same direction in $XY$- plane leads to zero steady magnetization current, see
Remark after (\ref{J_Heat}). Consequently, by imposing a twisting angle in
$XY$-plane of $\pi/2$ we choose a minimal nontrivial setup.

Transport properties in the $XXZ$ spin chain $H=\sum_{n}h_{n,n+1}$ are
governed by spin and energy current operators which are defined by lattice
continuity equations $\frac{d}{dt}\sigma_{n}^{z}=\hat{\jmath}_{n-1,n}%
-\hat{\jmath}_{n,n+1}$, \ $\frac{d}{dt}h_{n,n+1}=\hat{J}_{n}^{Q}-\hat{J}%
_{n+1}^{Q}$ where
\begin{equation}
\hat{\jmath}_{n,m}=2(\sigma_{n}^{x}\sigma_{m}^{y}-\sigma_{n}^{y}\sigma_{m}%
^{x}), \label{Jz}%
\end{equation}%
\begin{equation}
\hat{J}_{n}^{Q}=-\sigma_{n}^{z}\hat{\jmath}_{n-1,n+1}+\Delta(\hat{\jmath
}_{n-1,n}\sigma_{n+1}^{z}+\sigma_{n-1}^{z}\hat{\jmath}_{n,n+1}). \label{Q}%
\end{equation}

We show next that the magnetization current $j$ alternates its sign with the
system size, while the energy transport is switched off completely, if
$\kappa=\kappa^{\prime}$, despite the boundary gradients. Being locally
conserved quantities, in the stationary state the currents $\langle\hat
{\jmath}_{n,n+1}\rangle=j,$ $\langle\hat{J}_{n}^{Q}\rangle=J^{Q}$ are equal
across all the bonds.

For the following is essential to note, that for our choice of the Lindblad
dissipator (\ref{LindbladAction}), (\ref{LindbladOperators}), the system
(\ref{LME}) has a \textit{unique} nonequilibrium steady state. Its existence
and uniqueness for any coupling $\Gamma$ is guaranteed by the completeness of
the algebra, generated by the set of operators $\{H,V_{m},W_{m},V_{m}%
^{+},W_{m}^{+}\}$ under multiplication and addition \cite{EvansUniqueness},
and is verified straighforwardly as in \cite{ProsenUniqueness}.

\section{Symmetries of the Lindblad equation and restrictions on energy and
magnetization currents}

\label{sec::Symmetries of the Lindblad equation and restrictions on energy and magnetization currents}%

Let us denote by $\rho(N,\Delta,t)$ a time-dependent solution of the Lindblad
equation (\ref{LME}), (\ref{LindbladOperators}) for a system of $N$ sites, and
by $\rho(N,\Delta)=\lim_{t\rightarrow\infty}$ $\rho(N,\Delta,t)$ its steady
state solution, where $\Delta$ is the anisotropy of the $XXZ$ Hamiltonian
(\ref{Hamiltonian}). There exist transformations which map one LME solution to
another. Some of the transformations depend on the parity of $N$. For
\textit{even} $N$ , introducing $U=U^{\dagger}=%
{\textstyle\prod\limits_{n\text{ odd}}}
\otimes\sigma_{n}^{z}$, we find that $\rho(N,-\Delta,t)=U\rho^{\ast}%
(N,\Delta,t)U$ \ is a solution of the same Lindblad equation with the
anisotropy $-\Delta$. We denote by $\rho^{\ast}$ a complex conjugated matrix
in the basis where $\sigma^{z}$ is diagonal. However, if a nonequilibrium
steady state is unique, then a relation follows which relates the NESS for
positive and negative $\Delta$,
\begin{equation}
\rho(N,-\Delta)=U\rho^{\ast}(N,\Delta)U\text{ for }N\text{ even.}
\label{SymmetryEven}%
\end{equation}
For \textit{odd} $N$ we obtain another relation for NESS, introducing a
unitary transformation $\Sigma_{y}=(\sigma^{y})^{\otimes_{N}}$
\begin{equation}
\rho(N,-\Delta)=\Sigma_{y}U\rho^{\ast}(N,\Delta)U\Sigma_{y}\text{ for }N\text{
odd} \label{SymmetryOdd}%
\end{equation}

Another LME symmetry does not depend on parity of $N$ and maps a NESS\ onto
itself provided that an additional condition\ (\ref{SymmetryU-Uprime}) is
satisfied:
\begin{equation}
\rho(N,\Delta)=\Sigma_{x}U_{rot}R\rho(N,\Delta)RU_{rot}^{+}\Sigma_{x},
\label{SymmetryGlobal}%
\end{equation}
where $R(A\otimes B\otimes...\otimes C)=(C\otimes....\otimes B\otimes A)R$ is
a left-right reflection, and the diagonal matrix $U_{rot}=diag(1,i)^{\otimes
_{N}}$ is a rotation in $XY$ plane, $U_{rot}\sigma_{n}^{x}U_{rot}^{+}=$
$\sigma_{n}^{y}$, $U_{rot}\sigma_{n}^{y}U_{rot}^{+}=-\sigma_{n}^{x}$, and
$\Sigma_{x}=(\sigma^{x})^{\otimes_{N}}$. Note also that at the point
$\Delta=0$ symmetries (\ref{SymmetryEven}) and (\ref{SymmetryOdd}) map a
NESS\ onto itself, and are independent on (\ref{SymmetryGlobal}).

The relations (\ref{SymmetryEven}) -- (\ref{SymmetryGlobal}) impose
restrictions on the NESS density matrices and respectively on all observables
including the magnetic and energy currents $j=Tr(\rho\hat{\jmath}_{n,n+1})$
and $J^{Q}=Tr(\rho\hat{J}_{n}^{Q})$ where $\hat{\jmath}_{n,n+1}$ and $\hat
{J}_{n}^{Q}$ are given by (\ref{Jz}) and (\ref{Q}). We drop the subscripts
since in the steady state the currents $j,J^{Q}$ are equal across all bonds.
By virtue of the symmetry (\ref{SymmetryEven}), we obtain for $N$ even
\begin{align*}
j(-\Delta)  &  =Tr(\hat{\jmath}_{n,n+1}\rho(N,-\Delta))=\\
&  =Tr(\hat{\jmath}_{n,n+1}U\rho^{\ast}(N,\Delta)U)=\\
&  =-Tr(U\hat{\jmath}_{n,n+1}U\rho(N,\Delta))^{\ast}=\\
&  =Tr(\hat{\jmath}_{n,n+1}\rho(N,\Delta))=j(\Delta).
\end{align*}
In the last passage we used (i) $\hat{\jmath}_{n,n+1}^{\ast}=-$ $\hat{\jmath
}_{n,n+1}$ in the basis where $\sigma^{z}$ is diagonal (ii) $U\hat{\jmath
}_{n,n+1}U=-\hat{\jmath}_{n,n+1}$ (iii) the fact that an observable $j$ is a
real number. Analogously, for odd $N$, we obtain using (\ref{SymmetryOdd}),
that $j(-\Delta)=-j(-\Delta)$. So, we find that the magnetization current
changes its parity from even to odd as a function of the anisotropy $\Delta$:
\begin{align}
j(\Delta)  &  =j(-\Delta)\text{ for }N\text{ even,}\label{J_even}\\
j(\Delta)  &  =-j(-\Delta)\text{ \ for }N\text{ odd.} \label{J_odd}%
\end{align}
We argue below, see sec.\ref{subsec::Free fermion point. Arbitrary N}, that at
the free fermion point $\Delta=0$ the spin current vanishes always,
$j(\Delta=0)=0$. For odd $N$ it follows from (\ref{J_odd}), while for even $N$
it follows from a perturbative expansion. In addition, we make another
suprising observation: due to (\ref{J_even}), (\ref{J_odd}) the spin current
alternates its sign with the system size since by increasing a system size by
one unit $N\rightarrow N+1$ the current dependence on $\Delta$ changes from
even to odd or vice versa. For our case it amounts to
\begin{equation}
sign\text{ }j(\Delta,N)=(-1)^{N}\text{ for }\Delta\leq-1\text{,}
\label{CurrentSignAlternation}%
\end{equation}
because in the region $\left\vert \Delta\right\vert <1$ the sign of
$j(\Delta)$ can change depending on other parameters, see
sec.\ref{subsec::Arbitrary Delta. Finite size results}. The alternating
current effect is a consequence of an existence of different LME symmetries
for even and odd number of sites (\ref{SymmetryEven}), (\ref{SymmetryOdd}) and
as such is quite general. Indeed the derivation of (\ref{J_even}%
),(\ref{J_odd}) does not depend neither on coupling $\Gamma$ now on gradients
$\kappa$ and $\kappa^{\prime}$ as given by (\ref{LindbladOperators}). Also,
the phenomenon of alternating current does not rely on an integrability of the
$XXZ$ model. Indeed, an addition of a staggered anisotropy to the Hamiltonian
(\ref{Hamiltonian}) breaks integrability but the Eqs. (\ref{SymmetryEven}),
(\ref{SymmetryOdd}) remain valid. Moreover, even though our conclusion relied
on the validity of Eqs. (\ref{J_even}), (\ref{J_odd}), the alternating current
was also observed in a recent study of a driven $XXZ$ chain with an $XY$
boundary setup different from (\ref{LindbladOperators}), where one of two
symmetries (Eq(\ref{J_even})) was absent \cite{Lindblad2011}. We suggest that
the magnetization current sign alternation resulting from applying transverse
gradients can be a rather general phenomenon, which might be observable under
appropriate experimental conditions in artificially assembled nanomagnets
consisting of just a few \ atoms \cite{ScienceLoth11}.

Another consequence of the current sign alteration is that the spin
current\textit{ cannot be ballistic}. Indeed, for large $N$ we should not be
able to differentiate between even and odd $N$, so that the stationary
magnetization current $j$ must vanish in the thermodynamic limit:%
\begin{equation}
j\left\vert _{N\rightarrow\infty}\right.  =0\text{ for all }\Delta,\Gamma.
\label{CurrentNonBallistic}%
\end{equation}

This conclusion is a rather unexpected one, since in the $XXZ$ integrable
system, in the critical region $\left\vert \Delta\right\vert <1$ one usually
expects a ballistic current \cite{ZotosBallistic97}. The magnetization current
is also ballistic $\left\vert \Delta\right\vert <1$ for reservoirs creating
boundary gradients along the anisotropy axis \cite{ProsenMasur2011}.

Up to now we did not make use of the symmetry (\ref{SymmetryGlobal}). In fact,
it does not impose any further restrictions for spin current, while the energy
current changes its sign under the symmetry (\ref{SymmetryGlobal}) and must
therefore vanish
\begin{equation}
J^{Q}=0\text{ for all }N,\Delta,\Gamma. \label{J_Heat}%
\end{equation}
Consequently, the energy and spin currents decouple: the spin current can
flow, while the energy current is totally suppressed (\ref{J_Heat}). Vanishing
of the energy current is a consequence of a symmetric choice $\kappa
=\kappa^{\prime}$ (\ref{SymmetryU-Uprime}), and will be lifted if $\kappa
\neq\kappa^{\prime}$.

\textbf{Remark}. Note that if both boundary reservoirs were aligning boundary
spins along the same direction in $XY$-plane, the magnetization current will
be completely suppressed for all $\Delta$, despite the existence of boundary
gradients. E.g. for a choice of the set of Lindblad operators $V_{1}%
=\alpha(\sigma_{N}^{y}+i\sigma_{N}^{z}),V_{2}=0,$ $W_{1}=\beta(\sigma_{N}%
^{y}+i\sigma_{N}^{z}),$ $W_{2}=0$ in (\ref{LindbladAction}), which favours the
alignment of the boundary spins along the $X$ axis, the current suppression
follows from a symmetry $\rho=\Sigma_{x}\rho\Sigma_{x}$ of the NESS, since
$Tr(\hat{\jmath}_{n,n+1}\rho)=-Tr(\hat{\jmath}_{n,n+1}\Sigma_{x}\rho\Sigma
_{x})$. Thus, our choice (\ref{LindbladAction}) is the minimal nontrivial one,
to observe a nonzero magnetization current.

In the following we check our findings and analyze the model further by
computing exact NESS numerically for finite $\Gamma$ and analytically for
$\Gamma\rightarrow\infty$ using a perturbative expansion.

\section{Strongly driven $XXZ$ chain with $XY$ twist: numerical and analytical
study.}

\label{sec::Strongly driven XXZ chain : numerical and analytical study}

We performed studies of the LME\ for small system sizes $N$ in order to
support our findings (\ref{J_even}-\ref{J_Heat}), which should be valid for
all $N,\Delta$ and $\Gamma$. For simplicity we restrict here to the strongly
driven case $\kappa=\kappa^{\prime}=1$. Being a problem of exponential growing
complexity, the steady state for a finite $\Gamma$ is a very complicated
function of the parameters $\Gamma,\Delta$, for any $N>3$, and cannot reported
otherwise than in a graphical form. We solve the linear system of equations
which determine the full steady state, numerically, making use of the global
symmetry (\ref{SymmetryGlobal}) which decrease the number of uknowns by
roughly the factor of $1/2$.

\begin{figure}[ptbh]
\begin{center}
\includegraphics[width=0.4\textwidth]{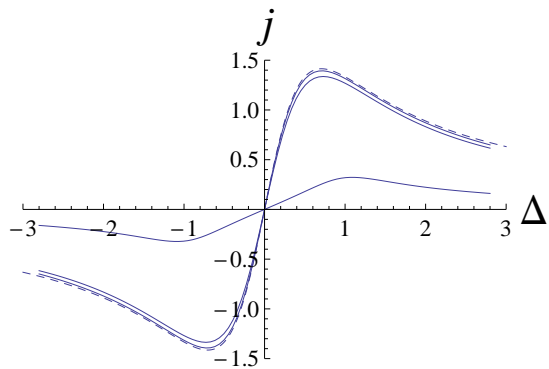}
\end{center}
\caption{ NESS spin current $j$ vs $\Delta$ for $N=3$ for different
$\Gamma=0.5,5,10$ (from bottom up). For finite $\Gamma$, the data are obtained
numerically. Dashed line marks $j(\Delta)$ in the limit $\Gamma\rightarrow
\infty$ from (\ref{J3})}%
\label{Fig_J3}%
\end{figure}

\begin{figure}[ptbh]
\begin{center}
\includegraphics[width=0.4\textwidth]{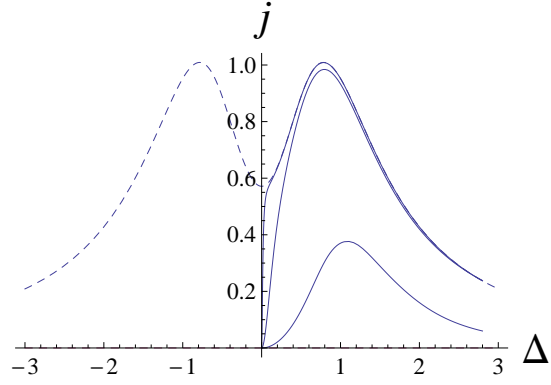}
\end{center}
\caption{ NESS spin current $j$ vs $\Delta$ for $N=4$ for different
$\Gamma=0.5,5,50$ (from bottom up). The $j(\Delta)$ curves continue
symmetrically in the negative $\Delta$ region. For finite $\Gamma$ the data
are obtained numerically. Dashed line marks $j(\Delta)$ in the limit
$\Gamma\rightarrow\infty$ from (\ref{J4}). }%
\label{Fig_J4}%
\end{figure}

For $N=3,4$ the magnetization currents as function of the anisotropy, for
various values of $\Gamma$ are reported in Fig.\ref{Fig_J3},\ref{Fig_J4}. As
expected from (\ref{J_even}),(\ref{J_odd}), the current is an odd (even)
function of $\Delta$ for $N=3(N=4)$. Interestingly, as the coupling $\Gamma$
increases, the current $j(\Gamma,\Delta)$ approaches quickly a limiting curve
drawn by a dashed line. For the dashed line, which corresponds to the the
limit of infinitely strong couplings $\Gamma\rightarrow\infty$, a simple
closed analytic form can be obtained. Here it is worthwhile to note that
Lindblad Master equation dynamics (\ref{LME}) with large effective coupling to
a reservoir $\Gamma\gg1$ is realizable within the experimental protocol
involving repeated periodic interactions of a boundary spin with a infinite
bath system of identical ancilla \cite{ClarkPriorMPA2010}.

To investigate the strong coupling limit $\Gamma\gg1$ we search for a
stationary solution of the Lindblad equation in the form of a perturbative
expansion
\begin{equation}
\rho_{NESS}(\Delta,\Gamma)=\sum_{k=0}^{\infty}\left(  \frac{1}{2\Gamma
}\right)  ^{k}\rho_{k}(\Delta). \label{PT_largeCouplings}%
\end{equation}

As stressed in the end of Sec.\ref{sec::The model}, the stationary solution
$\rho_{NESS}(\Delta,\Gamma)$ is unique. The zeroth order term of the expansion
(\ref{PT_largeCouplings}) $\rho_{0}=$ $\lim_{\Gamma\rightarrow\infty}%
\lim_{t\rightarrow\infty}\rho(\Gamma,\Delta,t)$ satisfies $%
\mathcal{L}%
_{LR}[\rho^{str}]=0$ (here and below we denote by $%
\mathcal{L}%
_{LR}=%
\mathcal{L}%
_{L}+%
\mathcal{L}%
_{R}$ the sum of the Lindblad actions in (\ref{LME})). This enforces a
factorized form
\begin{equation}
\rho_{0}=\rho_{L}\otimes\left(  \left(  \frac{I}{2}\right)  ^{\otimes_{N-2}%
}+M_{0}(\Delta)\right)  \otimes\rho_{R}, \label{InitialChoiceRo0}%
\end{equation}
where $\rho_{L}$ and $\rho_{R}$ are one-site density matrices (\ref{RoL}),
(\ref{RoR}) and $M_{0}$ is a matrix to be determined self-consistently later.
We shall drop $\Delta$- and $N$-dependence in $\rho$ for brevity of notations.
We separate the identity matrix $\left(  \frac{I}{2}\right)  ^{\otimes_{N-2}}$
for future convenience, so that $Tr(\rho_{0})=1$ and $M_{0}$ is traceless.

Inserting the (\ref{PT_largeCouplings}) into (\ref{LME}), and comparing the
orders of $\Gamma^{-k}$, we obtain recurrence relations%
\[
i[H,\rho_{k}]=\frac{1}{2}%
\mathcal{L}%
_{LR}\rho_{k+1}\text{ , \ }k=0,1,2...
\]
A formal solution of the above is $\rho_{k+1}=-2%
\mathcal{L}%
_{LR}^{-1}(Q_{k+1})$ where $Q_{k+1}=-i[H,\rho_{k}]$. Note however that the
operator $%
\mathcal{L}%
_{LR}$ has a nonempty kernel subspace, and is not invertible on the elements
from it. The kernel subspace $\ker(%
\mathcal{L}%
_{LR})$ consists of all matrices of type $\rho_{L}\otimes A\otimes\rho_{R}$
where $A$ is an arbitrary $2^{N-2}\times$ $2^{N-2}$ matrix. Therefore a
necessary and sufficient condition for $\rho_{k+1}$ to exist is to require a
null overlap
\begin{equation}
\lbrack H,\rho_{k}]\cap\ker(%
\mathcal{L}%
_{LR})=\varnothing\label{SecularConditionsGeneral}%
\end{equation}
We shall name (\ref{SecularConditionsGeneral}) the secular conditions. For our
choice of the Lindblad operator the secular condition are equivalent (see the
Appendix) to the requirement of a null partial trace
\begin{equation}
Tr_{1,N}([H,\rho_{k}])=0\text{, \ }k=0,1,2.... \label{SecularConditions}%
\end{equation}
The remaining missing ingredient of the perturbation theory is obtained by
noting that $\rho_{k+1}$ is defined up to an arbitrary element from $\ker(%
\mathcal{L}%
_{LR})$, so we have
\begin{equation}
\rho_{k+1}=2%
\mathcal{L}%
_{LR}^{-1}(i[H,\rho_{k}])+\rho_{L}\otimes M_{k+1}\otimes\rho_{R}\text{,
\ \ \ }k=0,1,2... \label{Recurrence}%
\end{equation}

Eqs (\ref{InitialChoiceRo0}), (\ref{Recurrence}) and (\ref{SecularConditions})
define a perturbation theory for the Lindblad equation (\ref{LME}). At each
order of the expansion, we must satisfy the secular conditions
(\ref{SecularConditionsGeneral}), before proceeding to the next order.
Construction of the inverse $%
\mathcal{L}%
_{LR}^{-1}$ is illustrated in the Appendix.

Applying the perturbation theory to our model, we find that the unknown
apriori matrices $M_{2k}(\Delta),M_{2k+1}(\Delta)$ are fully determined by
secular conditions (\ref{SecularConditions}) for $2k,2k+1$, as will be
illustrated further on an example. For any $\Delta\neq0$ the set of matrices
$\{M_{k}(\Delta)\}$ is nontrivial. The case of zero anisotropy $\Delta=0$
turns out to be special in that all $M_{k}\equiv0$ and is discussed separately below.

\subsection{\textbf{"Free fermion" point} $\Delta=0$. \textbf{Arbitrary }%
$N$\ }

\label{subsec::Free fermion point. Arbitrary N}

For $\Delta=0$ we can guess a zero approximation $\rho_{0}$ for arbitrary
$N$:
\begin{equation}
\rho_{0}=\rho_{L}\otimes\left(  \frac{I}{2}\right)  ^{\otimes_{N-2}}%
\otimes\rho_{R}. \label{R0_Delta0}%
\end{equation}
Comparing with (\ref{InitialChoiceRo0}), we see that $M_{0}=0$. All further
$M_{k}=0$ and all secular conditions (\ref{SecularConditions}) are trivially
satisfied at all orders. This fact was checked explicitly for small system
sizes $N\leq7$ and is conjectured to be true for arbitrary $N$. Consequently,
the general recurrence (\ref{Recurrence}) is reduced to $\rho_{k+1}=2%
\mathcal{L}%
_{LR}^{-1}(i[H,\rho_{k}])$ for all $k$. At the first order of expansion, we
obtain $i[H,\rho_{0}]=Q_{1}$, where
\begin{align*}
Q_{1}  &  =\frac{1}{2}\sigma^{z}\otimes\sigma^{x}\otimes\left(  \frac{I}%
{2}\right)  ^{\otimes_{N-3}}\otimes\rho_{R}\\
&  +\frac{1}{2}\rho_{L}\otimes\left(  \frac{I}{2}\right)  ^{\otimes_{N-3}%
}\otimes\sigma^{y}\otimes\sigma^{z}%
\end{align*}
which is easily inverted $\rho_{1}=2%
\mathcal{L}%
_{LR}^{-1}(Q_{1})=-Q_{1}$ and so on. At all orders we notice $Tr(\hat{\jmath
}\rho_{k})=0$, which allows to conjecture that magnetization current vanishes
for $\Delta=0$ in spite of twisting boundary gradients,
\begin{equation}
\left.  j\right\vert _{\Delta=0}=0\text{ \ \ \ for all }N\geq3\text{ and all
}\Gamma\text{.} \label{ZeroCurrDelta0}%
\end{equation}
Note that for odd $N$, the statement (\ref{ZeroCurrDelta0}) also follows from
(\ref{J_odd}).

For $N=3$, the perturbative expansion (\ref{PT_largeCouplings}) can be summed
up for all orders due to the fact that it closes at the third step: $\rho
_{3}=-4\rho_{1}$. Consequently, (\ref{PT_largeCouplings}) is rewritten as
\begin{equation}
\rho_{N=3,\Delta=0}(\Gamma)=\rho_{0}+\left(  \frac{1}{2\Gamma}\rho_{1}%
+\frac{1}{(2\Gamma)^{2}}\rho_{2}\right)  \sum\limits_{m=0}^{\infty}%
(-1)^{m}\frac{1}{\Gamma^{2m}}, \label{N3_De0_StrongCoupling}%
\end{equation}
where $\rho_{1}=-\frac{1}{2}(\sigma^{z}\otimes\sigma^{x}\otimes\rho_{R}%
+\rho_{L}\otimes\sigma^{y}\otimes\sigma^{z})$ and $\rho_{2}=\frac{1}{2}%
\sigma^{y}\otimes I\otimes\rho_{R}-\frac{1}{4}\sigma^{x}\otimes\sigma
^{z}\otimes\sigma^{z}+\frac{1}{4}\sigma^{z}\otimes\sigma^{z}\otimes\sigma
^{y}+\frac{1}{2}\rho_{L}\otimes I\otimes\sigma^{y}$. Summing up the series, we
obtain
\begin{equation}
\rho_{N=3,\Delta=0}(\Gamma)=\rho_{0}+\frac{\Gamma}{2}\frac{\rho_{1}}%
{\Gamma^{2}+1}+\frac{\rho_{2}}{4(\Gamma^{2}+1)}. \label{RhoDe0_N=3}%
\end{equation}
It can be verified straightforwardly that (\ref{RhoDe0_N=3}) is a stationary
solution of LME (\ref{LME}) for $N=3,\Delta=0$. The sum
(\ref{N3_De0_StrongCoupling}) \ converges for $\Gamma>1$. However, since the
LME solution is unique, the expression (\ref{RhoDe0_N=3}) is valid for any
$\Gamma$. In particular, for $\Gamma\rightarrow0$ we obtain the weak coupling
limit $\rho^{weak}(\Delta)=\lim_{\Gamma\rightarrow0}\lim_{t\rightarrow\infty
}\rho(\Gamma,\Delta,t)$, which is also unique by a continuity argument,
\[
\rho_{N=3,\Delta=0}(\Gamma\rightarrow0)=\left.  \rho^{weak}(\Delta)\right\vert
_{\Delta=0}=\rho_{0}+\frac{1}{4}\rho_{2}.
\]
As expected, $\left.  \rho^{weak}(\Delta)\right\vert _{\Delta=0}$ commutes
with the $XX0$ Hamitonian, $[\rho_{0}+\frac{1}{4}\rho_{2},H_{XX0}]=0$, and
corresponds to zero currents, $j(\Gamma\rightarrow0)=0$.

From the above example we see that perturbative series
(\ref{PT_largeCouplings}) are useful even outside of their radius of
convergence: if the perturbative series for $\rho_{NESS}$ or for a particular
observable $\langle\hat{f}\rangle=Tr(\hat{f}\rho_{NESS})$ can be summed, the
result of the summation is valid for all $\Gamma$ within its analyticity
range, see also (\ref{Delta1_N3}), (\ref{Delta1_N4}).

\subsection{ Exact NESS for small system sizes and arbitrary $\Delta.$}

\label{subsec::Arbitrary Delta. Finite size results}

For any nonzero $\Delta$, the matrices $M_{k}$ from (\ref{Recurrence}) are
nontrivial and are determined from secular conditions (\ref{SecularConditions}%
). The secular conditions for two consecutive orders $2m,2m+1$ reduce to a
system of linear equations which determine $M_{2m}$ and $M_{2m+1}$. \ In such
a way, the perturbative series (\ref{PT_largeCouplings}) is uniquely defined
at all orders.

Below we present analytic results for $M_{0}$ and small $N$ which define the
NESS through Eq.(\ref{InitialChoiceRo0}). $M_{0}$ and $M_{1}$ are obtained by
satisfying secular conditions (\ref{SecularConditions}) for $k=0$ and $k=1$.
The procedure is illustrated below for the simplest case $N=3,4$. For $N=3$,
the most general form of the matrices $M_{0}(\Delta)$, $M_{1}(\Delta)$, by
virtue of the symmetry (\ref{SymmetryGlobal}), is $M_{0}=b_{0}(\sigma
^{x}-\sigma^{y})$, $M_{1}=b_{1}(\sigma^{x}-\sigma^{y})$ where $b_{0},b_{1}$
are unknown constants. The secular conditions (\ref{SecularConditions}) for
$k=0$ do not give any constraints on $b_{0}$, while those for $k=1$ give two
nontrivial relations $b_{1}=0$ and $-\Delta+(1+2\Delta^{2})b_{0}=0$ from which
both $M_{0}$ and $M_{1}$ are determined. For $N=4$ each of the matrices
$M_{0}(\Delta),M_{1}(\Delta)$ compatible with the symmetry
(\ref{SymmetryGlobal}), contains $9$ unknowns, all fixed by the secular
conditions (\ref{SecularConditions}) for $k=0,1$, and so on. For $N=3,4$ we
obtain
\begin{equation}
\left.  M_{0}(\Delta)\right\vert _{N=3}=\frac{\Delta}{1+2\Delta^{2}}%
(\sigma^{x}-\sigma^{y}) \label{M0_N3}%
\end{equation}%
\begin{align}
\left.  s(\Delta)M_{0}(\Delta)\right\vert _{N=4}  &  =\left(  -\frac
{3\Delta^{2}}{2}+2\Delta^{4}\right)  \sigma^{z}\otimes\sigma^{z}+\left(
\frac{1}{2}-2\Delta^{2}\right)  \sigma^{x}\otimes\sigma^{y}\nonumber\\
&  -4\Delta^{2}\sigma^{y}\otimes\sigma^{x}+\left(  \frac{1}{2}+2\Delta
^{2}\right)  (\sigma^{x}\otimes I-I\otimes\sigma^{y})\label{M0_N4}\\
&  +\frac{\Delta(4\Delta^{2}+5)}{2}(I\otimes\sigma^{x}-\sigma^{y}\otimes
I)\nonumber\\
&  +2\Delta(\sigma^{x}\otimes\sigma^{x}+\sigma^{y}\otimes\sigma^{y}).\nonumber
\end{align}
where $s(\Delta)=7+6\Delta^{2}+8\Delta^{4}$. Matrices $M_{0}$ obey the
symmetries (\ref{SymmetryEven})--(\ref{SymmetryGlobal}). Steady state spin
currents $j_{0}(\Delta)=Tr(\rho_{0}\hat{\jmath})$ in the limit $\Gamma
\rightarrow\infty$ are
\begin{equation}
\left.  j_{0}(\Delta)\right\vert _{N=3}=\frac{4\Delta}{1+2\Delta^{2}},
\label{J3}%
\end{equation}%
\[
\left.  j_{0}(\Delta)\right\vert _{N=4}=\frac{4(1+4\Delta^{2})}{7+6\Delta
^{2}+8\Delta^{4}},
\]
while the energy current vanishes $J^{Q}=0$ as expected from (\ref{J_Heat}).
\ We note that the current for $N=4$ is nonzero at $\Delta=0$ which seems to
contradict our prediction (\ref{ZeroCurrDelta0}). The contradiction arises
because the limits $\Gamma^{-1}\rightarrow0$ and $\Delta\rightarrow0$ do not
commute. Indeed, we find that even orders of the perturbative expansion
$\rho_{2m}$ contain terms which become singular for $\Delta\rightarrow0$. The
leading singular term at the order $2m$ of the expansion is $Tr(\varepsilon
^{2m}\rho_{2m}\hat{\jmath})\sim(\varepsilon/\Delta)^{2m}$, where
$\varepsilon=\Gamma^{-1}$. Since the singularities become stronger with $m$,
for any given infinitesimal $\epsilon$ there will be a region $\left\vert
\Delta\right\vert \lesssim\epsilon$ where the expansion
(\ref{PT_largeCouplings}) will diverge. The origin behind the non-commuting
limits is the existence of an extra symmetry (\ref{J_even}) of the NESS at the
point $\Delta=0$. Similar situation is encountered e.g. in Kubo linear
response theory, where a nonergodicity at zero point in the momentum space
$\vec{q}=0$ is inflicted by an extra conservation law at this point
\cite{nonergodicity}. Reversing order of the limits $\epsilon\rightarrow0$ and
$\Delta\rightarrow0$ does yield $j(0)=0$ for any $\Gamma$, so that the current
expression for $N=4,\Gamma\rightarrow\infty$ has to be modified as
\begin{equation}
\left.  j_{0}(\Delta)\right\vert _{N=4}=%
\genfrac{\{}{.}{0pt}{}{\frac{4(1+4\Delta^{2})}{7+6\Delta^{2}+8\Delta^{4}%
},\text{ if }\Delta\neq0}{0,\text{ if }\Delta=0,}
\label{J4}%
\end{equation}
while $\left.  M_{0}\right\vert _{N=4,\Delta=0}=0$ as given by
(\ref{R0_Delta0}). For any finite $\Gamma$ the dependence $j(\Delta)$ is
smooth and does not contain any singularities, see Fig.\ref{Fig_J4}. We find
the noncommutativity of the limits $\epsilon\rightarrow0$, $\Delta
\rightarrow0$ for all even system sizes $N=2m\geq4$, see also (\ref{J6}).
Since for large $N$ one should not be able to differentiate between even and
odd system sizes, we expect that the $j_{0}(\Delta)$ for $\Delta\ll1$ vanishes
in the thermodynamic limit: $\lim_{m\rightarrow\infty}\left.  j_{0}(\Delta
\ll1)\right\vert _{N=2m}=0$.

\begin{figure}[ptbh]
\begin{center}
\subfigure[\label{fig:Ja}]
{\includegraphics[width=0.4\textwidth]{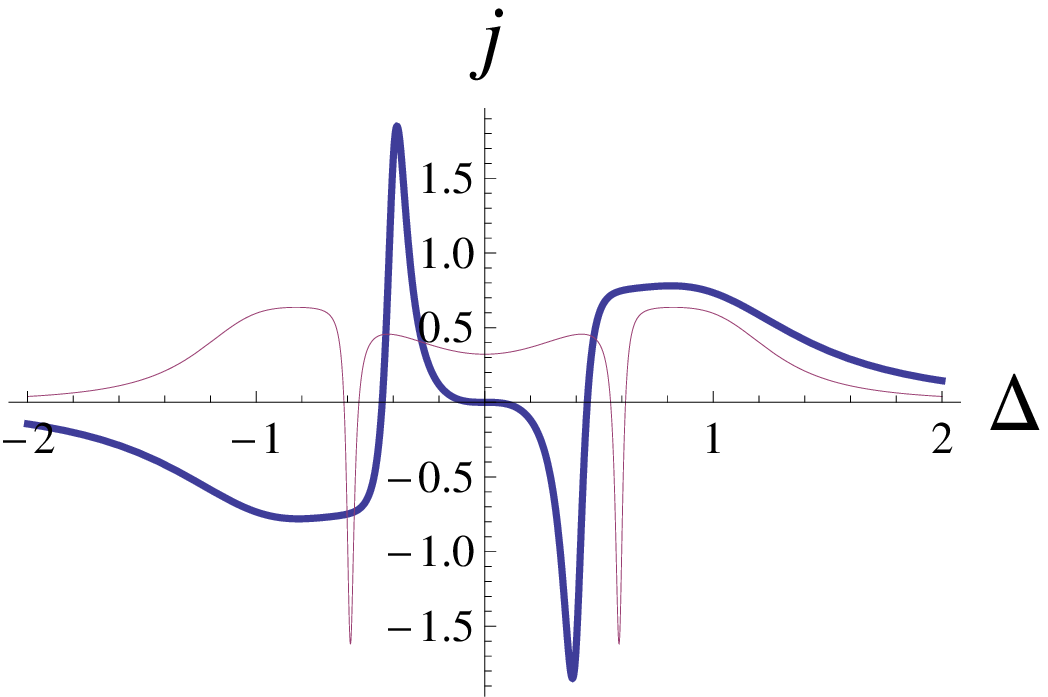}} \qquad
\subfigure[\label{fig:Jb}]{\includegraphics[width=0.4\textwidth]{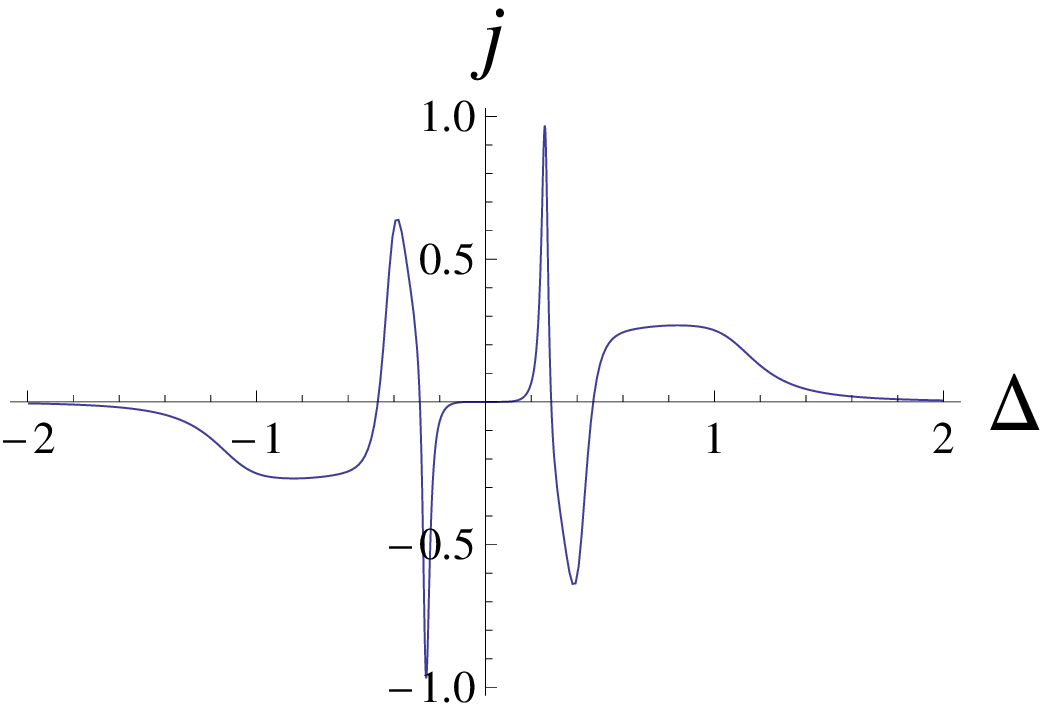}}
\end{center}
\caption{NESS spin current vs $\Delta$ for $N=5,6$ from Eqs(\ref{J5}%
),(\ref{J6}) given by thick and thin line respectively (Panel (a)), and for
$N=7$ (Panel (b)), all for $\Gamma\rightarrow\infty$.}%
\label{Fig_J5-7}%
\end{figure}

\begin{figure}[ptbh]
\begin{center}
\subfigure[\label{fig:XYa}]
{\includegraphics[width=0.4\textwidth]{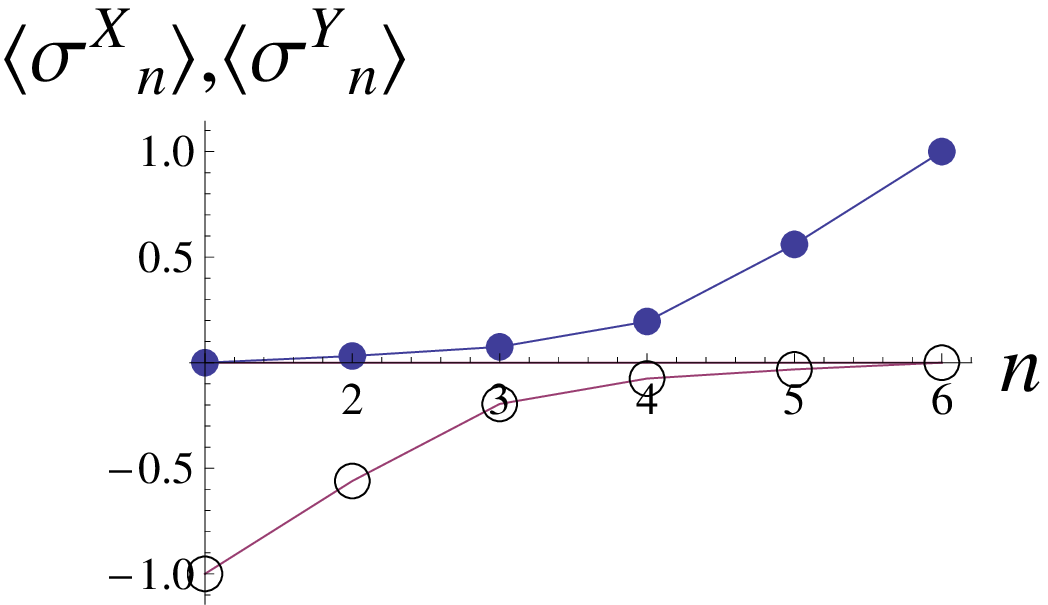}} \qquad
\subfigure[\label{fig:XYb}]
{\includegraphics[width=0.4\textwidth]{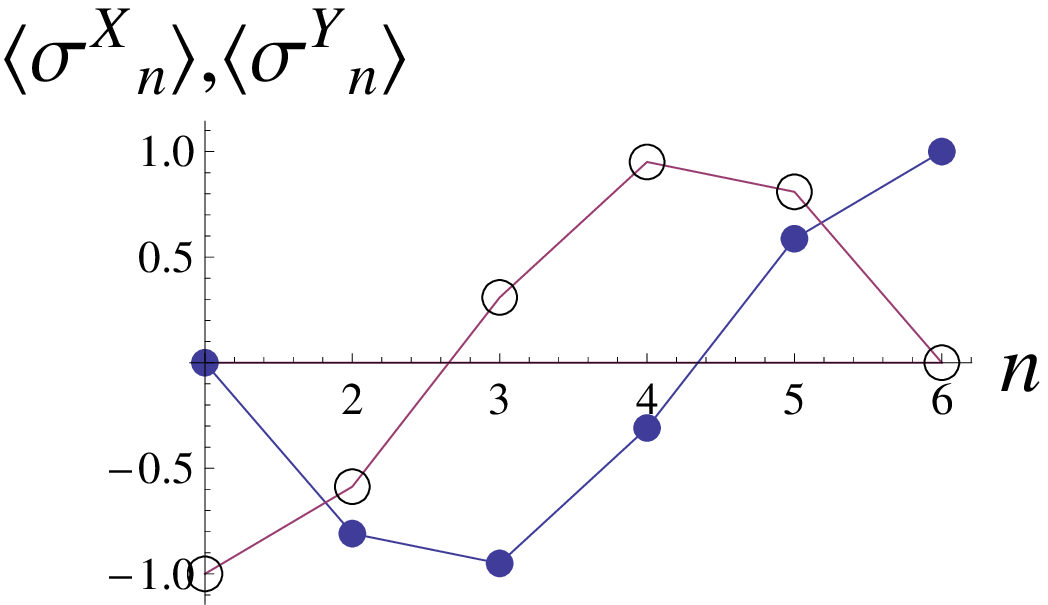}} \qquad
\subfigure[\label{fig:XYc}]{\includegraphics[width=0.4\textwidth]{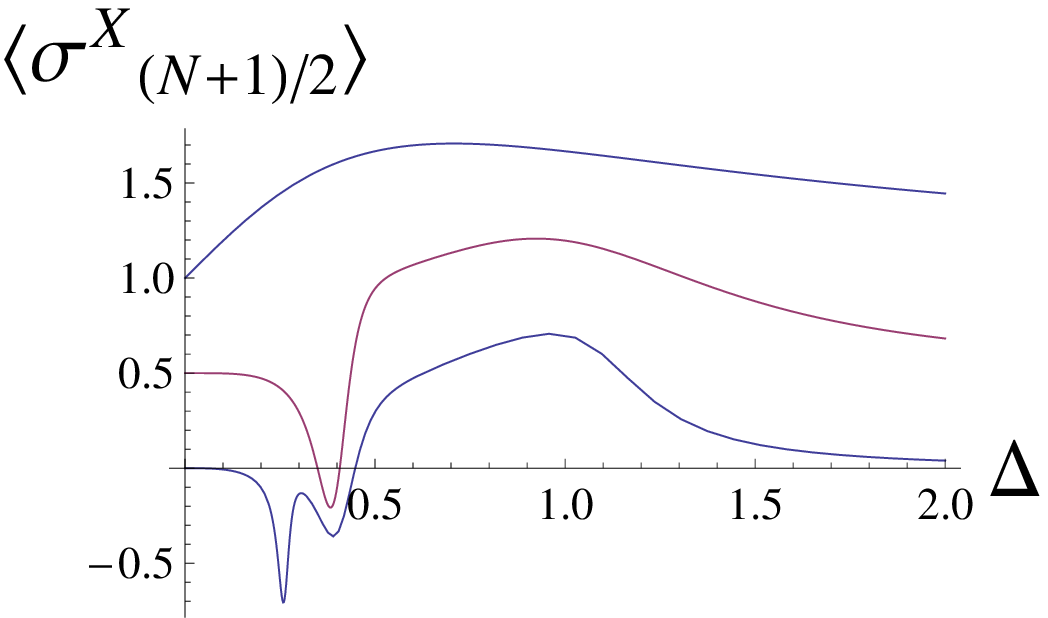}}
\end{center}
\caption{Typical $\langle\sigma_{n}^{X}\rangle$-profiles (filled circles) and
$\langle\sigma_{n}^{Y}\rangle$- profiles (open circles) outside the resonance
(Panel (a), $\Delta=1.8,N=6$) and inside the resonance (Panel (b),
$\Delta=0.5875,N=6$). Frequency tripling is clearly observed in the Panel (b).
Panel (c) shows dependence of the average spin projection $\langle\sigma
_{n}^{X}\rangle$ in the middle of the chain $n=(N+1)/2$ on $\Delta$, for
$N=3,5,7$ (from up to down). The curves for $N=3,5$ are shifted vertically by
$1$ and $0.5$ for a better view. }%
\label{Fig_XYprofiles}%
\end{figure}

The number of equations to solve (\ref{SecularConditions}) at each order of
perturbation theory grows exponentially with the system size $N$, although the
symmetry (\ref{SymmetryGlobal}) restricts the number of independent variables.
The matrices $M_{0}$ for $N\geq5$ are given by lengthy expressions and are not
reproduced here. Instead, we report NESS currents at $\Gamma\rightarrow\infty
$, given for $N=5,6$ by
\begin{equation}
\left.  j_{0}\right\vert _{N=5}=\frac{4\Delta^{3}(-3+12\Delta^{2}+2^{3}%
\Delta^{4}+2^{5}\Delta^{6})}{1-8\Delta^{2}-6\Delta^{4}+120\Delta^{6}%
+96\Delta^{8}-2^{6}\Delta^{4}+2^{7}\Delta^{12}} \label{J5}%
\end{equation}%
\begin{equation}
\left.  j_{0}\right\vert _{N=6}=%
\genfrac{\{}{.}{0pt}{}{\frac{4(12-3\Delta^{2}-220\Delta^{4}+48\Delta^{6}%
+2^{6}19\Delta^{8}-2^{8}5\Delta^{10}+2^{10}\Delta^{12})}{149-604\Delta
^{2}+316\Delta^{4}-912\Delta^{6}+4992\Delta^{8}-5760\Delta^{10}+15360\Delta
^{12}-2^{14}\Delta^{14}+2^{13}\Delta^{16}},\text{ if }\Delta\neq0.}{0,\text{
if }\Delta=0}
\label{J6}%
\end{equation}
The NESS currents $j_{0}$ for odd/even number of sites $N$ are even/odd
functions of $\Delta$ respectively, as expected from (\ref{J_even}),
(\ref{J_odd}) and depend rather nontrivially on the anisotropy $\Delta$, see
Fig.\ref{Fig_J5-7}. In particular for $N\geq5$ one observes sharp peaks in the
$j_{0}(\Delta)$ resembling resonances at which the current changes its sign
and amplitude. One gets an insight by comparing average magnetization profiles
$\langle\sigma_{n}^{x}\rangle$ and $\langle\sigma_{n}^{y}\rangle$ for values
of $\Delta$ inside and outside the resonance region. The average magnetization
at boundaries $n=1,N$ is fixed by the Lindblad reservoirs $\langle\sigma
_{1}^{x}\rangle\rightarrow0$, $\langle\sigma_{N}^{x}\rangle\rightarrow1$ and
$\langle\sigma_{1}^{y}\rangle\rightarrow1$, $\langle\sigma_{N}^{y}%
\rangle\rightarrow0$ so that the lowest interpolating Fourier mode has the
wavelength $\Lambda/a=4(N-1)$ where $a$ is the lattice constant. In the
resonance region we observe that $\Lambda%
\acute{}%
=\Lambda/3$ $\ $\ becomes the dominant Fourier mode, see
Fig.\ref{Fig_XYprofiles}(b). One can interpret this frequency tripling as a
triple increase of a twisting angle along the chain. Note, see
Fig.\ref{Fig_XYprofiles}(a,b) that there is a $\pi/2$ phase shift between the
dominant harmonics describing $\langle\sigma_{n}^{x}\rangle$ and
$\langle\sigma_{n}^{y}\rangle$ profiles. In particular, in the resonance
region, Fig.\ref{Fig_XYprofiles}(b), $\langle\sigma_{n}^{x}\rangle$
$\approx-\sin\phi(n-1),\langle\sigma_{n}^{y}\rangle$ $\approx-\cos\phi(n-1)$,
where $\phi=3\pi/10$. In the meanfield approximation the respective
magnetization current is $j_{MF}=2(\langle\sigma_{n}^{x}\rangle\langle
\sigma_{n+1}^{y}\rangle-\langle\sigma_{n}^{y}\rangle\langle\sigma_{n+1}%
^{x}\rangle)=-2\sin\phi\approx-1.61803$, which is amazingly close to the exact
current value at the resonance $j_{exact}=-1.61857..$. With the increase of
system size, one might expect an appearance of further resonances at
wavelengths $\Lambda%
\acute{}%
=\Lambda/5,\Lambda/7,...$ corresponding to larger twisting angles $\frac{5\pi
}{2},\frac{7\pi}{2}$ etc.. Why higher Fourier harmonics become stable under a
strong Lindblad action for narrow lacunae of $\Delta$, and how to predict a
location of the peaks remains an intriguing open question. As coupling
$\Gamma$ decreases, the resonances become smoother and then ultimately
disappear, but their precursors can still be seen for $\Gamma$ of the order of
the Heisenberg exchange interaction and larger (data not shown).

Examining analytic expressions for the current for small $N$, we can
extrapolate its behaviour for $\Delta\gg1$ and $0<\Delta\ll1$ \ and finite $N$
as
\begin{equation}
\left.  j\right\vert _{\Gamma\rightarrow\infty,\Delta\gg1}=O\left(  \frac
{1}{\Delta^{N-2}}\right)  \label{LimitJforDeltaLarge}%
\end{equation}%
\begin{equation}
\left.  j\right\vert _{\Gamma\rightarrow\infty,0<\Delta\ll1}=O((-1)^{\frac
{N+1}{2}}\Delta^{N-2})\text{ for odd }N \label{LimitJforDeltaSmall}%
\end{equation}
\ The Eq.(\ref{LimitJforDeltaLarge}) implies an exponential decay of the
current with the system size $N$ for fixed large $\Delta$, $\left.
j\right\vert _{\Gamma\rightarrow\infty,\Delta\gg1}=O(e^{-\alpha N})$ with
$\alpha=\ln\Delta$, and is generically expected from an insulating nature of
the $XXZ$ chain in the $\Delta\gg1$ Ising limit. Similar behaviour of the NESS
current at large $\Delta$ was found for driven chain with $Z$-polarized
boundary baths \cite{ProsenExact2011}. On the other hand,
Eq.(\ref{LimitJforDeltaSmall}) implies a complete flattening of the current
near $\Delta=0$ in the thermodynamic limit $N\rightarrow\infty$, which is a
new unusual feature. The flattening is already well seen in Fig.\ref{Fig_J5-7}
for $N=5,7$. Note that the results (\ref{LimitJforDeltaLarge}),
(\ref{LimitJforDeltaSmall}) would be very difficult to obtain by studying
perturbative series in $\Delta$ or in $1/\Delta$: in both cases $\Delta\ll1$
and $1/\Delta\ll1$ the first nonzero term will appear only in the order $N-2$
of the expansion!

\begin{figure}[ptbh]
\begin{center}
\includegraphics[width=0.4\textwidth]{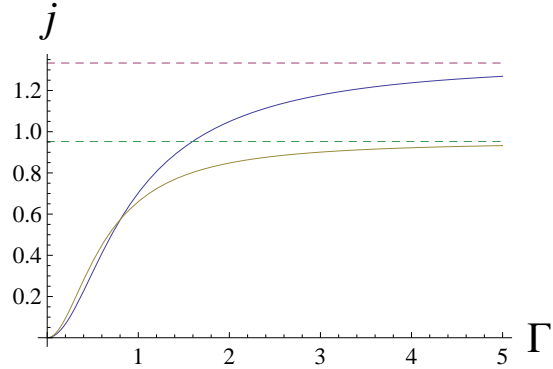}
\end{center}
\caption{ NESS spin current $j$ vs $\Gamma$ for $N=3,4$ (thick and thin line,
respectively) for isotropic Heisenberg model $\Delta=1$, given by
(\ref{Delta1_N3}),(\ref{Delta1_N4}). Dashed lines indicate limiting values of
the current for $\Gamma\rightarrow\infty$.}%
\label{Fig_JversusGA}%
\end{figure}

\subsection{Generic $\Gamma$ and arbitrary driving $\kappa$}

\label{subsec::Generic Gamma and arbitrary driving}

The analytic results derived in previous subsection concerned the limits of
$\Gamma\rightarrow\infty$ and strong driving $\kappa=\kappa^{\prime}=1$. What
can we say about generic $\Gamma$ and $\kappa$? Examining perturbative series
(\ref{PT_largeCouplings}) for various $N$ and $\Delta$ we notice that nonzero
contributions to NESS\ current appear only at even orders of the expansion
$\Gamma^{-2n}$. If the NESS current $j(\Gamma,\Delta)$ is an analytic function
of $\Gamma$, then we can expect the current to be an even function of $\Gamma
$. In particular, for small $\Gamma\ll1$ one expects the the current to behave
as $j(\Gamma\ll1)=O(\Gamma^{2})$, since the current must vanish at zero
coupling $\Gamma=0$. Our expectations are confirmed by exact expressions for
the current as function of $\Gamma$, obtained for special case of isotropic
$XXX$ model, for which one can sum up the series (\ref{PT_largeCouplings}) and
obtain
\begin{align}
\left.  j(\Gamma)\right\vert _{N=3,\Delta=1}  &  =\frac{16\Gamma^{2}%
(3+\Gamma^{2})}{27+52\Gamma^{2}+12\Gamma^{4}}\label{Delta1_N3}\\
\left.  j(\Gamma)\right\vert _{N=4,\Delta=1}  &  =\frac{8\Gamma^{2}%
(27+46\Gamma^{2}+10\Gamma^{4})}{3(27+136\Gamma^{2}+144\Gamma^{4}+28\Gamma
^{6})}. \label{Delta1_N4}%
\end{align}
Indeed the above expressions contain only even $\Gamma$ orders and
$j(\Gamma\ll1)=O(\Gamma^{2})$, see also Fig.\ref{Fig_JversusGA}. Numerically,
we observe $j(\Gamma\ll1)=O(\Gamma^{2})$ for other $\Delta$ values as well.
Generically, we find the current amplitude to be a smooth function of $\Gamma$
for a fixed value of the anisotropy $\Delta$. For large couplings, it
asymptotically approaches its limiting value $j_{0}(\Delta)$ at $\Gamma
\rightarrow\infty$. Interestingly, the magnetization current is not suppressed
for large couplings $\Gamma$, as it happens for $Z$-polarized Lindblad baths
\cite{ProsenExact2011},\cite{BenentiPRB2009} due to a quantum Zeno effect.
Indeed for our choice of the bath (\ref{LindbladOperators}) and large
coupling, the quantum Zeno effect "freezes" $\sigma^{y}$ and $\sigma^{x}$ spin
components at the boundaries, while hoppings in the $\sigma^{z}$ component,
contributing to the magnetization current, may still occur.

Finally, we investigated the current dependence on the boundary gradient
$\kappa$, which can be manipulated by a choice of the amplitudes of the
Lindblad operators in (\ref{LindbladOperators}). By Fourier law, we may expect
the current to be proportional to $\kappa$, for small gradients $\kappa$, and
introduce a conductivity tensor $\chi_{\alpha\beta}$ as a proportionality
coefficient between the current $j\equiv j^{z}$ and the infinitesimal boundary
gradient $\langle\delta\sigma^{\beta}\rangle/N$ in the spin component $\beta$,
$j^{\alpha}=\chi_{\alpha\beta}\langle\frac{\delta\sigma^{\beta}}{N}\rangle$.

With our perturbative method we construct iteratively an expansion
(\ref{PT_largeCouplings}) for a NESS for an arbitrary amplitude of the
boundary gradient $\kappa$ from (\ref{LindbladOperators}). We observe that the
current is a rather nontrivial function of $\kappa$, and in particular that it
vanishes as $\kappa^{2}$ or faster for small $\kappa$. Consequently,
$\lim_{\kappa\rightarrow0}j(\kappa)/\kappa=0$ so that the off-diagonal
elements of the conductivity tensor $\chi_{zx},\chi_{zy}$ vanish. Note that
diagonal elements of the conductivity tensor are nonzero \cite{BenentiPRB2009}%
. Numerically, we find that the leading order of expansion of $j(\kappa
)/\kappa$ for small $\kappa$ depends on $N$ and $\Delta$ in a nontrivial way.
More details will be given elsewhere.

Finally, note that our model shows a behaviour fundamentally different from
that with usually considered boundary driving along the $Z$-axis, referred to
below as a "scalar" setup. Just to mention two crucial differences: the scalar
setup leads to a ballistic current in the critical gapless region of the $XXZ$
model $\Delta<1$ \cite{ProsenExact2011}, while in the $XY$-driven chain the
alternating signs phenomenon (\ref{SymmetryEven}-\ref{CurrentSignAlternation})
rules out the possibility of a ballistic current. In the "scalar" setup the
current at strong coupling is suppressed due to a Zeno effect, $j_{\Gamma
\rightarrow\infty}\rightarrow0$, while in the $XY$-driven chain of finite size
the spin current in the limit $\Gamma\rightarrow\infty$ remains finite.

\section{Conclusions}

\label{sec::Conclusions}

We considered an open $XXZ$ spin 1/2 chain with a twisting in $XY$- plane,
imposed by boundary gradients. Symmetries of the Lindblad equation were found
which impose drastic restrictions on the steady state reduced density matrix,
and various observables. For our boundary setup the NESS energy current
vanishes, and the magnetization current alternates its sign with system size,
which rules out the possibility of a ballistic current. We argued that the
current sign alternation under trasverse gradients is a robust phenomenon in
quantum transport which does not rely on integrability. The qualitative
difference in quantum transport in even- and odd-sized systems adds to a list
of similar phenomena observed in quantum diffusion in periodic chains
\cite{TorreEvenOdd}. To support our findings, we constructed a nonequilibrium
steady state solution of the Lindblad Master equation in the form of a
perturbative expansion in orders of $\Gamma^{-1}$, and calculated the
zeroth-order term analytically for small system sizes, by solving a set of
linear equations which guarantee the self-consistency of the expansion, the
secular conditions (\ref{SecularConditions}). We find that, for $\Delta\neq0$,
the current remains finite even in the limit of infinitely strong effective coupling.

Further on, for large couplings we find a nontrivial dependence of the
magnetization current on the anysotropy $j(\Delta)$, and observed sharp peaks
in $j(\Delta)$ inside the critical gapless region of the $XXZ$ model
$\Delta<1$, the origin of which was attributed to an appearance of higher
Fourier harmonics in magnetization profiles. We observe an anomalous
flattening of the current near $\Delta=0$ point for odd $N$ and
noncommutativity of limits $\Delta\rightarrow0$ and $\Gamma^{-1}\rightarrow0$
for even $N$. For weak driving $\kappa\ll1$ and small system sizes, we find
$\lim_{\kappa\rightarrow0}j(\kappa)/\kappa=0$ for all values of $\Delta$,
signalizing subdiffusive current in the direction trasverse to $XY$ boundary gradients.

\textbf{Acknowledgements }The author thanks N. Plakida, R. Vaja, M. G. Pini,
M. Salerno and R. Moessner for stimulating discussions. Support from the
italian MIUR through PRIN 20083C8XFZ initiative is acknowledged.

\appendix

\section{Inverse of the Lindblad dissipators and secular conditions.}

$%
\mathcal{L}%
_{L}$ and $%
\mathcal{L}%
_{R}$ are linear super-operators acting on a matrix $\rho$ as defined by
(\ref{LindbladAction}). In our case, each super-operator act locally on a
single qubit only. We find the eigen-basis $\{\phi_{R}^{\alpha}\}_{\alpha
=1}^{4}$ of $\frac{1}{2}%
\mathcal{L}%
_{R}\phi_{R}^{\alpha}=\lambda_{\alpha}\phi_{R}^{\alpha}$ in the form $\phi
_{R}=\{\rho_{R},\sigma^{x},\sigma^{y},\sigma^{z}\},$ where $\sigma^{x}%
,\sigma^{y},\sigma^{z}$ are Pauli matrices and $\rho_{R}=\frac{1}{2}\left(
I+\kappa\sigma^{x}\right)  $. The respective eigenvalues are $\{\lambda
_{\alpha}\}=\{0,2,1,1\}$. Analogously, the eigen-basis and eigenvalues of the
eigenproblem $\frac{1}{2}%
\mathcal{L}%
_{L}\phi_{L}^{\beta}=\mu_{\beta}\phi_{L}^{\beta}$ are $\phi_{L}=\{\rho
_{L},\sigma^{x},\sigma^{y},\sigma^{z}\}$ and $\{\mu_{\beta}\}=\{0,1,2,1\}$,
where $\rho_{L}=\frac{1}{2}\left(  I-\kappa\sigma^{y}\right)  $. Since the
basises $\phi_{R}$ and $\phi_{L}$ are complete, any matrix $F$ acting in the
appropriate Hilbert space can be expanded as
\begin{equation}
F=\sum\limits_{\alpha=1}^{4}\sum\limits_{\beta=1}^{4}\phi_{L}^{\beta}\otimes
F_{\beta\alpha}\otimes\phi_{R}^{\alpha} \label{ro_expansion}%
\end{equation}
in the unique way. Indeed, let us introduce complementary basises $\psi
_{R}=\{I,\frac{1}{2}(\sigma^{x}-\kappa I),\frac{1}{2}\sigma^{y},\frac{1}%
{2}\sigma^{z}\}$ and \ $\psi_{L}=\{I,\frac{1}{2}\sigma^{x},\frac{1}{2}%
(\sigma^{y}+\kappa I),\frac{1}{2}\sigma^{z}\}$, trace-orthonormal to the
$\phi_{R},\phi_{L}$ respectively, $Tr(\psi_{R}^{\gamma}\phi_{R}^{\alpha
})=\delta_{\alpha\gamma}$, $Tr(\psi_{L}^{\gamma}\phi_{L}^{\beta}%
)=\delta_{\beta\gamma}$. Then, the coefficients of the expansion
(\ref{ro_expansion}) are given by $F_{\beta\alpha}=Tr_{1,N}((\psi_{L}^{\beta
}\otimes I^{\otimes_{N-1}})F(I^{\otimes_{N-1}}\otimes\psi_{R}^{\alpha}))$. On
the other hand, in terms of the expansion (\ref{ro_expansion}) the
superoperator inverse $(%
\mathcal{L}%
_{L}+%
\mathcal{L}%
_{R})^{-1}$ is simply%
\begin{equation}
-2(%
\mathcal{L}%
_{L}+%
\mathcal{L}%
_{R})^{-1}F=\sum\limits_{\alpha,\beta}\frac{1}{\lambda_{\alpha}+\mu_{\beta}%
}\phi_{L}^{\beta}\otimes F_{\beta\alpha}\otimes\phi_{R}^{\alpha}.
\label{LindbladInversion}%
\end{equation}
Note however that the above sum contains a singular term with $\alpha=\beta
=1$, because $\lambda_{1}+\mu_{1}=0.$ To eliminate the singularity, one must
require $F_{11}=Tr_{1,N}F=0$, which produces the secular conditions
(\ref{SecularConditions}).


\begin{thebibliography}{99}
\bibitem {LargeCouplingsRemark}Within the repeated interaction scheme
\cite{ClarkPriorMPA2010}, \cite{RepeatedInteractionScheme}, the LME
(\ref{LME}) describes a system evolution where coupling to reservoir is
rescaled as $\Gamma/\sqrt{\tau}$ where $\tau$ is a time interval between the
two successive interactions with the ancilla bath.
\bibitem {TwistingRemark}If $\kappa=\kappa^{\prime}<1$, the situation is more
subtle since then the target states $\rho_{L}$ and $\rho_{R}$ are mixed
\end{thebibliography}
\end{document}